\documentclass[preprint,12pt]{elsarticle}

\usepackage{amsthm,amsmath,amssymb}
\usepackage{bm}
\usepackage[utf8]{inputenc} 
\usepackage{graphicx}
\graphicspath{{./Figs/}}
\usepackage{cancel}
\usepackage{booktabs}
\usepackage{hyperref}
\usepackage{siunitx}
\newcommand*\diff{\mathop{}\!\mathrm{d}}
\usepackage{upgreek}
\usepackage{graphicx}
\usepackage{subfigure}
\usepackage{color}
\usepackage[normalem]{ulem} % strike through
\newcommand{\Replace}[2]{\bgroup\noindent\textcolor{red}{\xout{#1} #2}\egroup\ignorespacesafterend}
\newcommand{\Delete} [1]{\bgroup\noindent\textcolor{red}{\xout{#1}}\egroup\ignorespacesafterend}
\newcommand{\Insert} [1]{\bgroup\noindent\textcolor{}{#1}\egroup\ignorespacesafterend}
\newcommand{\Comment}[1]{\definecolor{Mygray}{gray}{0.50}\bgroup\color{Mygray}\noindent#1\egroup\ignorespacesafterend}
\newcommand \Michael [1]{\bgroup\noindent[\textcolor{blue}{\textbf{Michael}: #1}]\egroup\ignorespacesafterend}
\newcommand \Stefan  [1]{\bgroup\noindent[\textcolor{blue}{\textbf{Stefan}: #1}]\egroup\ignorespacesafterend}
\DeclareMathAlphabet{\Ibb}{U}{msb}{m}{n}

\newcommand{\BA}{{\boldsymbol{\mathnormal A}}}

\newcommand{\BC}{{\boldsymbol{\mathnormal C}}}

\newcommand{\BL}{{\boldsymbol{\mathnormal L}}}

\newcommand{\BP}{{\boldsymbol{\mathnormal P}}}

\newcommand{\BI}{{\boldsymbol{\mathnormal I}}}
\newcommand{\BH}{{\boldsymbol{\mathnormal H}}}
\newcommand{\Bxi}    {\ensuremath{\boldsymbol\xi}}

\newcommand{\Bomega  }{\ensuremath{\boldsymbol\omega}}

\newcommand{\Bb}{{\boldsymbol{\mathnormal b}}}

\newcommand{\Be}{{\boldsymbol{\mathnormal e}}}
\newcommand{\Bf}{{\boldsymbol{\mathnormal f}}}

\newcommand{\Br}{{\boldsymbol{\mathnormal r}}}

\newcommand{\Bu}{{\boldsymbol{\mathnormal u}}}

\newcommand{\Bx}{{\boldsymbol{\mathnormal x}}}

\newcommand{\Beps    }{\ensuremath{\boldsymbol\epsilon}}

\newcommand \MZ [1] {\bgroup\noindent[\textcolor{blue}{\textbf{MZ}: #1}]\egroup\ignorespacesafterend}

\journal{Forces in Mechanics}

\begin{document}

	%%% Start of article front matter
	\begin{frontmatter}
		
		\title{	An energetically consistent surface correction method for bond-based peridynamics}
		\author[1,2]        
		{Jonas Ritter}
		\author[1,2]        
		{Shucheta Shegufta}
		\author[3]
		{Paul Steinmann}
		\author[1]
		{Michael Zaiser}

		\address[1]{Department of Materials Science, WW8-Materials Simulation, Friedrich-Alexander Universität Erlangen-Nürnberg (FAU), Dr.-Mack-str. 77, 90762 Fürth, Germany} 
		\address[2]{Central Institute for Scientific Computing (ZISC), Friedrich-Alexander Universität Erlangen-Nürnberg (FAU), Martensstrasse 5a, 91058 Erlangen, Germany}     
		\address[3]{Department of Mechanical Engineering,  Applied Mechanics, Friedrich-Alexander Universität Erlangen-Nürnberg (FAU), Egerlandstr. 5, 91058 Erlangen, Germany}

		%%%%%%%%%%%%%%%%%%%%%%%%%%%%%%%%%%%%%%%%%%%%%%
		%%                                          %%
		%% The Abstract begins here                 %%
		%%                                          %%
		%% Please refer to the Instructions for     %%
		%% authors on http://www.biomedcentral.com  %%
		%% and include the section headings         %%
		%% accordingly for your article type.       %%
		%%                                          %%
		%%%%%%%%%%%%%%%%%%%%%%%%%%%%%%%%%%%%%%%%%%%%%%

		\begin{abstract} % abstract
		A novel surface correction method is proposed for bond based peridynamics which ensures energy consistency with a classical reference body for general affine deformations. This method is validated for simple geometries and then applied to a typical surface-dominated problem, namely the indentation of a surface in the shallow to moderate-depth regime. 
		\end{abstract}

		
	  \end{frontmatter}

\section[Introduction]{Introduction}
\label{sec:1}

Peridynamics is a nonlocal formulation of continuum mechanics that was introduced by Silling \cite{silling2000reformulation} but which has close relations to Eringen's nonlocal elasticity theory \cite{eringen1983differential} and earlier work dating back to the 1960s (see e.g. Kröner \cite{kroner1967elasticity}) while its treatment of fracture problems shows close analogies to damage mechanics \cite{krajcinovic1996damage}. In its simplest form, so-called bond-based peridynamics, each material point is envisaged to interact with all material points within a finite domain, the so-called horizon, through central 'bond' forces whose magnitudes are proportional to the bond elongation times a phenomenological bond strength. Local fracture can then be simply described by setting the strength of a bond to zero upon fulfilment of a failure criterion. No specific traction boundary conditions are needed at surfaces or crack surfaces, which renders the method most useful for geometrically complex fracture problems such as fragmentation \cite{lai2015peridynamics} or fracture of highly porous media \cite{chen2019peridynamic,shen2021peridynamic}. 

A generic feature of peridynamics is the so-called surface effect: near-surface regions behave elastically softer than the bulk of the material. This is not always a desirable feature, as the peridynamic surface softening may misrepresent the behavior of actual materials. Here we present a new energy-based method to correct this surface effect if needed. We give a brief introduction into bond-base peridynamics in \autoref{sec:2}, including a discussion of the origin of the surface effect. We then present our correction scheme in \autoref{sec:3} and give examples of its application in \autoref{sec:4} before concluding in \autoref{sec:5}. 

\section{Theoretical Background}
\label{sec:2} 

For completeness of presentation, we give a brief overview of the bond based peridynamics model which we consider in the following; for original reference, see \cite{silling2010peridynamic}, our presentation follows mainly Ref. \cite{shen2021peridynamic}. We characterize the deformation of a $D$ dimensional continuous body ${\cal B}$ of density $\rho(\Bx)$ by the displacement field $\Bu(\Bx)$ where $\Bx$ are material coordinates.  The force balance equation for the point $\Bx$ is written in the form
\begin{equation}
	\rho(\Bx) \ddot{\Bu}(\Bx) = \int_{{\cal H}_{\Bx}} \Bf(\Bx,\Bx') \diff \Bx' + \Bb(\Bx) ,
\end{equation}
where $\Bf(\Bx,\Bx')$ is the pair force between $\Bx'$ and $\Bx$, $\Bb$ is a body force field, and interactions are restricted to the so-called horizon ${\cal H}_{\Bx}$ which we take to be a $D$ dimensional sphere of radius $\delta$ around $\Bx$, $(|\Bx - \Bx^*| \le \delta) \; \forall \; \Bx^* \in {\cal H}_{\Bx}$. 

The pair force is specified constitutively. Considering linear elasticity and small deformations, we write the pair force as
\begin{equation}
	\Bf(\Bx,\Bx') = \hat{\Bf}(\Bx,\Bx') + \hat{\Bf}(\Bx',\Bx) \quad,\quad
	\hat{\Bf}(\Bx,\Bx') = \frac{1}{2} \BC(\Bx,\Bxi)[\Bu(\Bx')-\Bu(\Bx)].
\end{equation}
Here $\Bxi = \Bx'-\Bx$. The micro-modulus tensor $\BC(\Bx,\Bxi)$ is of the form
\begin{equation}
	\BC(\Bx,\Bxi)= \frac{c(\Bx,\Bxi)}{\xi}[\Be_{\Bxi}\otimes \Be_{\Bxi}]
\end{equation}
where $\xi = |\Bxi|$, $\Be_{\Bxi} = \Bxi/\xi$, and the bond strength function $c(\Bx,\Bxi)$ is specified in such a manner as to achieve strain energy equivalence between the peridynamic continuum and a reference continuum which we take to be isotropic linear elastic with Poisson number $\nu = 1/4$ in 3D and $\nu = 1/3$ in 2D, such as to match the pair force interaction. The strain energy of a body ${\cal B}$ can for the peridynamic continuum be written as
\begin{equation}
	E = \int_{\cal B} W_{\rm p}(\Bx) \diff \Bx
\end{equation}
where the strain energy density $W_{\rm p}(\Bx)$ associated with the point $\Bx$ is given by an integral over the horizon  ${\cal H}_{\Bx}$:
\begin{equation}
	W_{\rm p}(\Bx) = \frac{1}{4} \int_{{\cal H}_{\Bx}} \frac{c(\Bx,\Bxi)}{\xi}  [\Be_{\Bxi}.[\Bu(\Bx')-\Bu(\Bx)]]^2\diff \Bx'.
	\label{eq:strainEnergyDensity}
\end{equation}
This must be matched with the standard elastic strain energy density of the reference continuum, $W_{\rm e} = 1/2 \Beps : \BH :  \Beps$ where $\Beps$ is the symmetrized gradient of $\Bu$ and $\BH$ is Hooke's tensor for the isotropic reference material. The result depends on spatial dimensionality as well as on a further constitutive choice as the radial dependency of $c(\Bx,\Bxi)$ needs to be specified. Popular choices in the literature are to assume $c = c_0(\Bx)$ independent of $\Bxi$, or to use a 'conical' form $c = c_0(\Bx) [1-\xi/\delta]$. Either way, in the bulk of a spatially homogeneous material, this matching leads to a bond strength $c_{\rm b}(\xi)$ that does not explicitly depend on $\Bx$, and that is direction-independent as required for material isotropy. We note that similar approaches may be used for other 'flavors' of peridynamics, e.g., energy matching for affine deformations has been used to relate the material parameters of the recently proposed continuum-kinematics-inspired peridynamics \cite{javili2019continuum} to the Lame parameters of an isotropic linear elastic continuum description, see Ekiz et al. \cite{ekiz2022relationships,ekiz2022two}. This leads to similar surface effects as described here for a bond-based model. 

For a material point $\Bx$ of distance less than $\delta$ from the surface $\partial {\cal B}$ of the deforming body, we denote the surface-truncated horizon by $\bar{\cal H}_{\Bx}$. For such a point, using the same bond strength values as in the bulk leads to a  energetic mismatch between the peridynamic and reference energy densities, as the parts of ${\cal H}_{\Bx}$ outside the material body do not contribute to the energy density in $\Bx$. Thus, the peridynamic energy density near the surface is reduced. This effect, if uncorrected, leads to a characteristic softening of the regions within a distance of $\delta$ from the surface. 

The surface effect is an intrinsic feature of peridynamics (and one it shares with other theories such as nonlocal elasticity) but it may not always be a desirable one. We first note that there exist a range of materials where the peridynamic surface effect can be exploited to represent real material behavior, e.g. in deformation of disordered cellular structures where surface softening may result from incompleteness of cells that intersect the sample surface \cite{liebenstein2018size}. In such situations, by appropriate calibration of the horizon size $\delta$ and the bond strength function $c(\xi)$, the peridynamic surface effect can be matched to the actual material response. However, this is not always the case. In deformation problems at surfaces, such as in indentation or wear, or in surface dominated systems such as highly porous solids, unphysical surface softening may  significantly skew the results. Since it can nevertheless be desirable to use peridynamics for such problems, e.g. because of its convenience in dealing with multiple cracking phenomena, strategies for mitigating the surface effect are needed. While convergence to classical behavior can always be ensured by using very small values of $\delta$, this may be prohibitive for reasons of numerical cost. Thus some kind of surface correction is required.

\section{A new energy based surface correction method}
\label{sec:3}

A wide range of approaches have been proposed in the literature to correct the peridynamic surface effect. An overview and critical discussion was given by Le and Bobaru \cite{le2018surface}. Here we focus exclusively on approaches that are based on the idea of adjusting the bond strengths in the near-surface region such as to restore, at least approximately, the energy equivalence between a peridynamic system and a classical reference system. 

The method described in the literature \cite{madenci2014peridynamic,le2018surface,shen2021peridynamic} may go back to Oterkus \cite{oterkus2010peridynamic} and can be summarized as follows: One evaluates displacements for a geometrically similar reference medium which is loaded under a range of different boundary conditions, typically by applying homogeneous surface loads along the $x,y,z$ axes, to obtain reference displacement fields $u^{[x],[y],[z]}(\Br)$. Inserting these displacements into \autoref{eq:strainEnergyDensity} and using the bulk bond strength $c_{\rm b}(\xi)$ yields an energy density $W_{\rm p}$ which, for near surface values of $\Bx$, is less than the reference energy density $W_{\rm e}$. One then defines  correction factors $h^{[x],[y],[z]}(\Bx) = W_{\rm e}({u^{[x],[y],[z]}})/W_{\rm p}({u^{[x],[y],[z]}})$. Since the correction factors depend, via the boundary loads, on the reference displacement fields $u^{[x],[y],[z]}(\Br)$, each different type of boundary condition produces a different correction factor and some sort of democratic compromise is evaluated iteratively. The entire procedure leaves many open questions since it is not easy to see how correction factors evaluated based on homogeneous boundary loads along different axes, either individually or through iterative compromise, could conceivably capture the local behavior during, say, an indentation experiment close to an embedded void. In support of this approach, Madenci and Oterkus \cite{madenci2014peridynamic} argue that {\em "since the presence of free surfaces is problem dependent, it is impractical to resolve this issue analytically"}. We shall demonstrate that this assertion is incorrect and that an analytical solution is, at least within some reasonable approximations, not too difficult.

We start from the energy equivalence relation for a bulk material point, \autoref{eq:strainEnergyDensity}: 
\begin{equation}
	W_{\rm p}(\Bx) = \frac{1}{4} \int_{{\cal H}_{\Bx}} \frac{c_{\rm b}(\xi)}{\xi} [\Be_{\Bxi}.[\Bu(\Bx')-\Bu(\Bx)]]^2\diff \Bx' = \frac{1}{2}\Beps(\Bx): \BH : \Beps(\Bx).
	\label{straine2}
\end{equation}
We now consider the case of a weakly varying strain field which is near-constant over the horizon such that with reasonable accuracy we may set $\Bu(\Bx') = \Bu(\Bx) + \BL(\Bx):\Bxi$ where $\BL = \Beps + \Bomega$ is the deformation gradient tensor. Since the rotation part $\Bomega$ does not contribute to the energy density, we obtain
\begin{equation}
	\frac{1}{2} \int_{{\cal H}_{\Bx}} \frac{c_{\rm b}(\xi)}{\xi}  [\Be_{\Bxi}.\Beps(\Bx).\Bxi]^2\diff \Bx' = \Beps(\Bx): \BH : \Beps(\Bx).
	\label{straine2a}
\end{equation}
We introduce the rank-four projection tensor $\BP_{\Bxi} = \Be_{\Bxi}\otimes\Be_{\Bxi}\otimes\Be_{\Bxi}\otimes\Be_{\Bxi}$ which we use to write $[\Be_{\Bxi}.\Beps.\Bxi]^2 = \xi^2 [\Beps:\BP_{\xi}:\Beps]$. This allows us to re-phrase the bulk energy equivalence relation as 
\begin{equation}
	\Beps(\Bx): \BI_{{\cal H}_{\Bx}} : \Beps(\Bx) = \Beps(\Bx): \BH : \Beps(\Bx)
    \quad,\quad \BI_{{\cal H}_{\Bx}} = \int_{{\cal H}_{\Bx}} \BP_{\Bxi} c_{\rm b}(\xi) \xi \diff \Bxi
\end{equation}
where we have replaced the integration over $\Bx'$ by an integration over $\Bxi$. 
Since this relationship must hold for any strain field, we conclude that the energy equivalence relationship in the bulk reduces to $\BI_{{\cal H}_{\Bx}} = \BH$ (note that we might as well drop the subscript $\Bx$ since the energy equivalence relationship for a bulk point is not position specific). This is indeed the relationship commonly used to parameterize the function $c_{\rm b}(\xi)$ for the bulk material. 

We now apply exactly the same approach to a material point with incomplete horizon $\bar{\cal H}_{\Bx}$ where, to achieve energy equivalence, the bond strengths $c_{\rm b}$ must be modified in a spatially and directionally specific manner, $c_{\rm b}(\xi) \to c(\Bx,\Bxi)$. Energy equivalence leads to the requirement
\begin{equation}
	\BH = \BI_{\bar{\cal H}_{\Bx}} \quad,\quad
	\BI_{\bar{\cal H}_{\Bx}} = \int_{\bar{\cal H}_{\Bx}} \BP_{\Bxi} c(\Bx,\Bxi) \xi \diff \Bxi.
\end{equation}
It follows that 
\begin{equation}
	\int_{{\cal H}_{\Bx}} \BP_{\Bxi} c_{\rm b}(\xi) \xi \diff \Bxi  =: \int_{\bar{\cal H}_{\Bx}} \BP_{\Bxi} c(\Bx,\Bxi) \xi  \diff \Bxi.
	\label{straine4}
\end{equation}
To proceed, it is convenient to express the integrals in spherical coordinates. Using that the tensor $\BP_{\Bxi}$ depends only on angular coordinates, we get
\begin{equation}
	\BI_{{\cal H}_{\Bx}} = 
	\int_{\Omega}  \BP_{\Bxi} 
	\int_0^{\delta} c_{\rm b}(\xi) \xi^D \diff \xi \diff \Omega = 
	\BI_{\bar{\cal H}_{\Bx}} 
	= 	\int_{\Omega}  \BP_{\Bxi}  
	\int_{0}^{d_{\Bx,\Bxi}} c(\Bx,\Bxi)  \xi^D \diff \xi \diff \Omega\\
	\label{eq:EnergyEquivalence}
\end{equation}
where $\Omega$ is the unit sphere in $D$ dimensions. The upper boundary $d_{\Bx,\Bxi}$ of the $\xi$ integration for the truncated horizon is obtained as illustrated in \autoref{fig:correction}: 
\begin{figure}[tbh]
	\centering
	\hbox{}
	\includegraphics[width=0.7\textwidth]{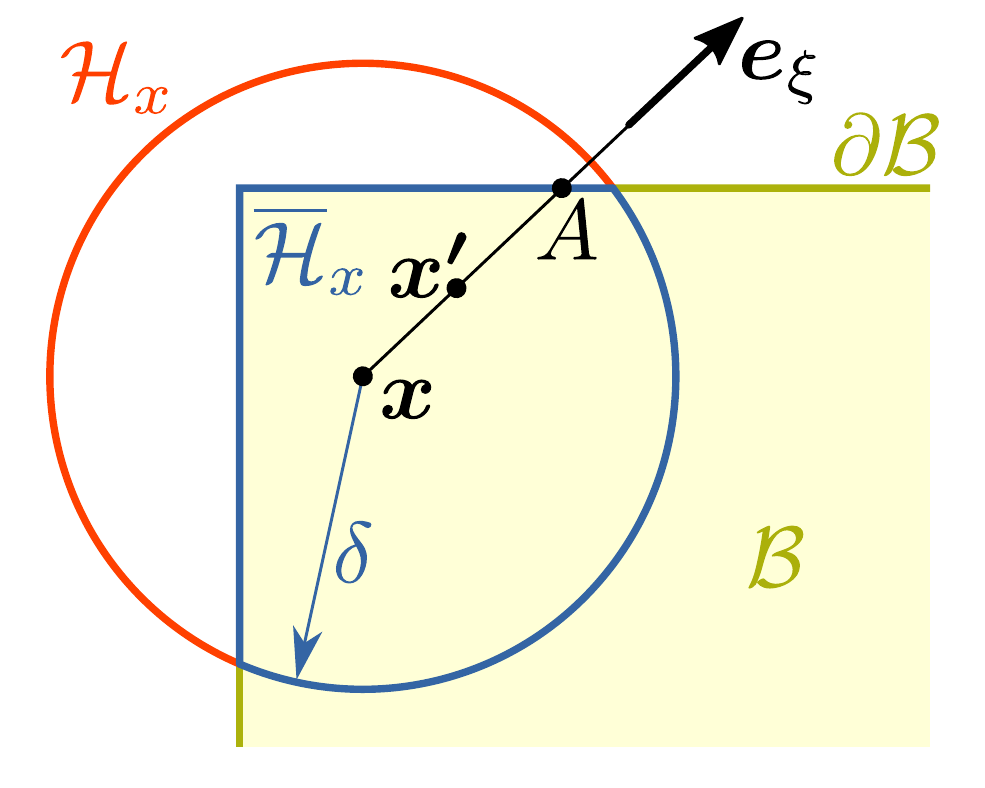}\hfill
	\caption{\label{fig:correction}
		Evaluation of the correction factor for bonds of direction $\Be_{\xi}$ connecting to a point $\Bx'$ near the surface of the body ${\cal B}$. The truncated horizon of $\Bx$ is $\bar{\cal H}_{\Bx}$ and the full horizon is ${\cal H}_{\Bx}$.  }
\end{figure}
The line through $\Bx$ with direction $\Be_{\xi}$ intersects the body surface $\partial {\cal B}$ in the point $\BA = \Bx + a \Be_{\xi}$. We then set
\begin{equation}
	d_{\Bx,\Bxi} = \min(a,\delta)
\end{equation}
i.e., in the example of \autoref{fig:correction}, $d_{\Bx,\Bxi} = a = \left\vert \Bx-\BA \right\vert$. The energy equivalence relation, \autoref{eq:EnergyEquivalence}, can then be fulfilled by equating, on both sides, the factors which multiply $\BP$ in the angular integration. This leads to the relationship
\begin{equation}
	 c(\Bx,\Bxi) = c_{\rm b}(\xi) \phi(\Bx,\Be_{\xi}) 
\end{equation}
where the direction dependent bond strength correction factor $ \phi(\Bx,\Be_{\xi})$ is given by
\begin{equation}
	\displaystyle
	\phi(\Bx,\Bxi) = \frac{\int_0^{\delta} c_{\rm b}(\xi) \xi^{D} \diff \xi}{\int_{0}^{d_{\Bx,\Bxi}} c_{\rm b}(\xi)  \xi^D \diff \xi} = \left(\frac{\delta}{d_{\Bx,\Bxi}}\right)^{D+1}.
	\label{eq:correctionFactor}
\end{equation}

\section{Applications}
\label{sec:4}

When numerically testing our correction scheme, some complications arise due to the peculiarities of peridynamics numerical implementations. Peridynamics is well known to converge to the classical continuum in the limit $\delta \to 0$ ($\delta$ convergence), and finite values of $\delta$ imply in general deviations from the classical continuum. However, peridynamics 
with finite $\delta$ is still a continuum theory whose numerical implementation requires some form of discretization, e.g. in form of finite elements (see e.g \cite{macek2007peridynamics}) or using meshfree methods \cite{silling2005meshfree, seleson2016convergence}. Such discretization introduces a discretization length $\Delta$, e.g in the form of finite element sizes or spacings of collocation points, and the discretized numerical solution converges to the exact peridynamic solution in the limit $m = \Delta/\delta = 0$ ($m$ convergence). Such convergence does {\em not} ensure numerical equivalence with the classical continuum, which is in general only achieved in the dual limit $\delta \to 0$, $m \to 0$. This is the simple reason why peridynamics, in order to achieve numerical equivalence with a classical continuum model, in general needs a very significantly larger number of degrees of freedom to describe a given problem than, for instance, a FEM model of equal accuracy.

Here we are interested in pragmatic approaches to mitigate this problem, and to reduce peridynamics simulations to manageable numerical cost. We therefore consider situations where $m$ convergence is approximately fulfilled, but $\delta$ convergence is not (i.e., the horizon is not small as compared to the specimen dimensions or other relevant scales of the deformation problem). We focus on two-dimensional problems and use, for numerical implementation of the peridynamics models,  a particle-based meshfree discretization scheme where a set of nodes with fixed associated volumes in the undeformed reference configuration and corresponding fixed masses is used to discretize the system. Such meshfree approaches are widely used in peridynamics because of their implementation simplicity and moderate computational cost. We will assume the nodes to form a simple cubic lattice of lattice constant $\Delta$. This means that we are, strictly speaking, simulating a medium of cubic synmmetry which however for $m \to 0$ converges towards an isotropic medium. We benchmark the performance of the peridynamic models against finite element calcuations which use a regular grid of square elements with the same lattice constant $\Delta$. We consider static solutions of various boundary value problems, which for the peridynamics models are evaluated using a quasi-static solver of the open-source code Peridigm \cite{Parks2012}.

\subsection{Simple geometries}

We consider two 2D problems, namely the tensile deformation of a rectangular sheet with force boundary conditions such as to induce a purely uni-axial stress state, and the displacement-controlled tensile deformation of a square sheet with boundaries clamped in tensile direction such as to induce multi-axial stresses and stress concentrations in the specimen corners. 

\subsubsection{Rectangular sheet, simple tension}

\begin{figure}[t]
	\centering
	\hbox{}
	\includegraphics[width=0.63\textwidth]{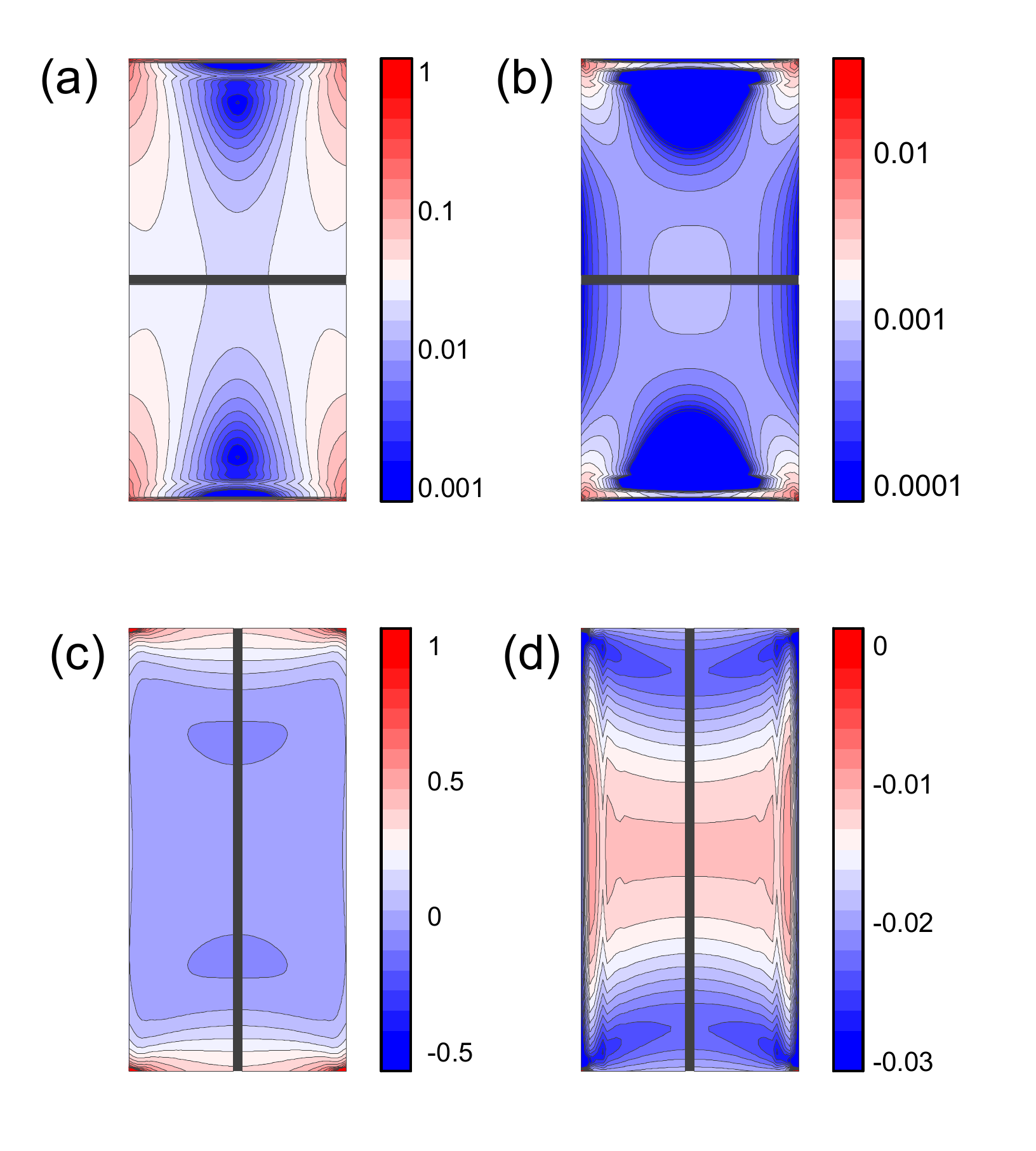}\hfill
	\caption{\label{fig:ResultFreeBoundary}
		Relative errors of the displacement fields for uni-axial deformation of a free standing sheet of size $\SI{50}{\mm} \times \SI{100}{\mm}$ with constant boundary loads imposed in $y$ (vertical) direction on the surfaces $y = \pm 50$; errors are evaluated relative to the analytical solution; (a) $\Delta u_y/u_y$, bond based peridynamics without surface correction, (b) $\Delta u_y/u_y$, bond based peridynamics with surface correction, (c) $\Delta u_x/u_x$, bond based peridynamics without surface correction, (d) $\Delta u_x/u_x$, bond based peridynamics with surface correction; the black bars represent nodes where for symmetry reasons both the PD result and the analytical displacement are zero such that a relative error is mathematically undefined.}
\end{figure}

This is a benchmark problem considered in the comparative study of Le and Bobaru \cite{le2018surface}, hence, the results can be used to comparatively evaluate the performance of the present boundary correction method. Following Le and Bobaru, we consider a square sheet of size $\SI{50}{\mm} \times \SI{100}{\mm}$  and elastic modulus $\SI{1}{\GPa}$. $51 \times 101$ discretization nodes are located on a regular square grid of lattice constant $\SI{1}{\mm}$. A homogeneously distributed force is applied in $y$ direction to the end surface nodes located at $y = \pm \SI{50}{\mm}$ such as to create a surface traction of $\SI{1}{\MPa}$. We use a horizon of radius $\delta = \SI{5}{\mm}$, hence $m = 1/5$ as also considered by Le and Bobaru. Differing from Le and Bobaru, who use a 'conical' parameterization where $C(\xi) = C_0[1-\xi/\delta]$ decreases linearly over the horizon, we parameterize the model assuming the micro-modulus to be constant over the horizon, $C(\xi) = C_0$. As we shall see, this choice exacerbates surface effects and therefore provides a more critical test of our correction method. 

We evaluate the performance of our surface correction method by comparing with the analytical solution of the problem which, to avoid confounding effects due to the finite value of $m$, we parameterize with the bulk values of elastic modulus and Poisson number for our discretization scheme. 
Results are shown in \autoref{fig:ResultFreeBoundary} which shows relative errors of the displacement fields in $x$ and $y$ directions, for uncorrected as well as for surface corrected simulations. The corresponding maximum errors are compiled in \autoref{tab:ResultFreeBoundary}.

We observe that the errors for the uncorrected PD simulation are about a factor 3 higher than those reported for the same set-up by Le and Bobaru \cite{le2018surface}. This is a direct consequence of the different way we parameterize the micro-modulus: a constant micro-modulus, as used here, tends to enhance surface effects as compared to the 'conical' scheme used by Le and Bobaru. Even so, the present correction scheme outperforms all surface correction methods considered by Le and Bobaru not only in relative, but in absolute terms. The single exception is the fictitious nodes method studied by Le and Bobaru which indeed is exact (but for issues of numerical accuracy) for the simple boundary conditions and boundary geometries considered in the present problem. 

\begin{table}[hbt]
	\centering
	\begin{tabular}{l l l}
		\toprule
		Simulation method & Maximum error, $u_x$ &  Maximum error, $u_y$\\ 
		\cmidrule(r){1-1}\cmidrule(lr){2-2}\cmidrule(l){3-3}
		Bond based PD, uncorrected & \SI{82}{\percent} & \SI{291}{\%} \\
		Bond based PD, corrected & \SI{2.6}{\%} & \SI{3.2}{\%} \\
		\bottomrule
	\end{tabular}
	\caption{\label{tab:ResultFreeBoundary}Maximum displacement errors, rectangular sheet with tensile force on end surfaces, tensile direction is parallel to $y$ axis.}
\end{table}

\subsubsection{Square sheet with clamped boundaries in tensile direction}

Throughout the following simulations, space is measured in units of the horizon $\delta$ and we set $m = 1/6$ which provides a good compromise between numerical accuracy (deviations from exact isotropy are less than \SI{2}{\%}) and efficiency. Since no analytical solution is available, static 2D reference FEM calculations are performed for an isotropic elastic medium with Young's modulus $E = \SI{1}{\GPa}$ and Poisson number $\nu = 1/3$, assuming plane stress conditions, and peridynamics simulations are parameterized such that their bulk behavior matches these properties. 

\begin{figure}[tbh]
	\centering
	\hbox{}
	\includegraphics[width=\textwidth]{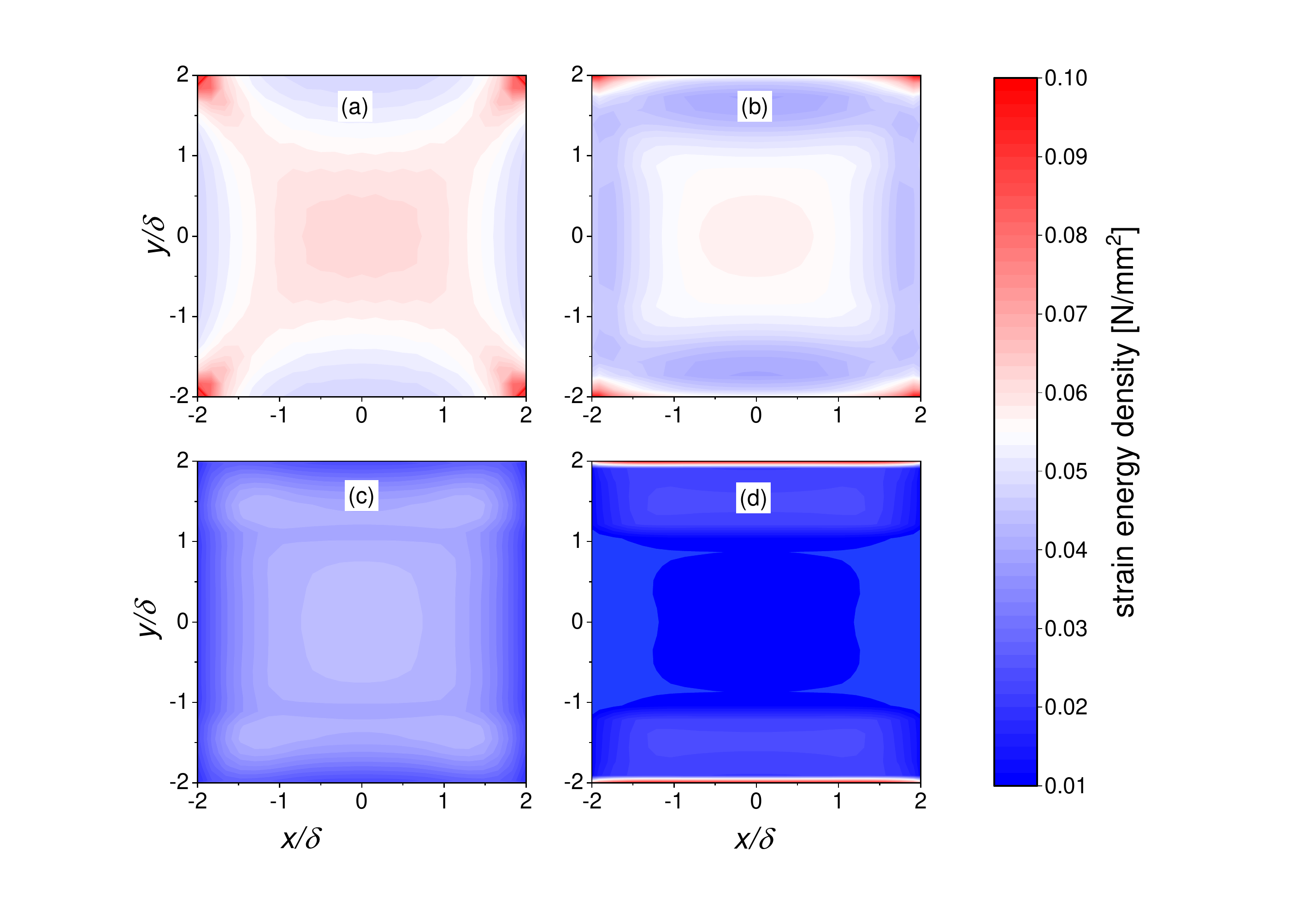}\hfill
	\caption{\label{fig:ResultClampedBoundary}Spatial patterns of strain energy density for the simulation methods in Table 1, (a) FEM reference, (b) corrected PD,  (c) PD with virtual nodes on constrained surfaces and uncorrected side surfaces, (d) uncorrected PD.}
\end{figure}

In our next test, we modify the boundary conditions with respect to the previous example. We simulate tensile deformation of a square sheet of edge length $L = 4 \delta$ with edges aligned along the $x$ and $y$ axes of a Cartesian coordinate system. The edges parallel to the $x$ axis are free, whereas the edges parallel to the $y$ axis are fully constrained. The upper and lower edges ($y=\pm 2 \delta$) are displaced in $y$ direction by $u_y = \pm 2\delta/100$, respectively, such as to induce an average axial strain of \SI{1}{\%}. Compared to the previous problem, this problem has three additional complexities: (i) surface effects are exacerbated since only 1/4 of the nodal points has full horizon, for all other points the horizon intersects the sample boundary; (ii) we have mixed boundary conditions; (ii) the imposed boundary conditions lead, for a classical continuum, to concentrations of stress, strain, and elastic energy in the corners of the sheet. Besides our FEM reference calculations, we consider the following variants of a peridynamic simulation:

\begin{enumerate}
	\item A PD simulation using a grid of $25 \times 25$ nodes, such that the outermost nodes are located directly at the surface. Surface displacments are imposed by moving the outermost points $y_0 = \pm  2\delta$) outward in $y$ direction by $y_0/100$ while constraining them at their original $x$ values, such as to induce an average axial strain of \SI{1}{\%}. Bond strengths are not corrected. 
	\item Often, displacement boundary conditions are imposed by using a layer of virtual nodes. We use two layers of thickness $\delta$ (i.e., six nodes in $y$ direction) outside the top and bottom surfaces. Here we use a grid of $24 \times [6 + 24 + 6]$ nodes with two buffer layers, such that the surface is located midway between the outermost points of the actual sample and the buffer layers. All grid points in the buffer layers are displaced rigidly in $y$ direction by $u_y = \pm 2\delta[1+\delta/12]/100$ while being constrained at their original $x$ values.   
	\item As (1) but with bond strengths corrected using \autoref{eq:correctionFactor}.
	\item As (2) with bond strengths corrected using \autoref{eq:correctionFactor} only on the side surfaces (note that no correction is needed on the top and bottom surface, where the correction is effected by the layers of virtual nodes)
\end{enumerate}
Macroscopic results are compiled in \autoref{tab:ResultClampedBoundary}. It is seen that the uncorrected peridynamic solution produces a tensile stress which amounts to only \SI{40}{\%} of the  FEM reference. This is not unexpected, since for a system width of $4 \delta$, only 1/4 of the sheet area is unaffected by peridynamic surface effects. The present correction scheme reduces the overall error to about \SI{1}{\%}, which is comparable to the numerical errors arising from the FEM discretization. Using a rigid buffer layer of virtual nodes at the constrained surfaces also produces some improvement, as the macroscopic stiffness of the system is approximately doubled relative to the uncorrected simulation and reaches about \SI{82}{\%} of the reference value. On the other hand, additional corrections on the side surfaces appear to be of minor importance and lead only to an additional stiffening by about \SI{4}{\%} of the reference value. 
\begin{table}[hbt]
	\centering
	\begin{tabular}{p{8cm} l}
		\toprule
		Simulation method & Tensile stress\\ 
		\cmidrule(r){1-1} \cmidrule(lr){2-2}
		1. FEM reference & \SI{10.32}{\MPa}\\
		\midrule
		2. PD uncorrected & \SI{4.17}{\MPa}\\
		3. PD corrected & \SI{10.21}{\MPa}\\ 
		4. PD, virtual nodes on constrained surfaces uncorrected side surfaces & \SI{8.43}{\MPa}\\
		5. PD, virtual nodes on constrained surfaces and corrected side surfaces & \SI{8.64}{\MPa}\\
		\bottomrule
	\end{tabular}
	\caption{\label{tab:ResultClampedBoundary}Tensile stress, square sheet with clamped end surfaces, global strain \SI{1}{\%}.}
\end{table}
We now turn to the spatial distribution of the elastic energy density as shown in \autoref{fig:ResultClampedBoundary}. The energy corrected peridynamics calculation produces results that are, in general, in good agreement with the FEM reference data shown in \autoref{fig:ResultClampedBoundary}(a). Notably, as can be seen by comparing \autoref{fig:ResultClampedBoundary}(a) and (b), the enhanced elastic energy in the corners and the reduction of elastic energy in a zone around the edges of the sample are captured. 

If, instead, the correction of the surface effect is performed by adding a horizon-wide layer of virtual nodes outside the constrained surfaces as shown in \autoref{fig:ResultClampedBoundary}(c), several interesting artefacts arise. The regions of reduced energy density below the constrained top and bottom surfaces are here replaced by a layer of {\em enhanced} energy of width $\delta$; this area of the sample has direct interaction with the virtual nodes of the displaced buffer layer. It may also be noted that part of the elastic energy density must be associated to the virtual nodes and is thus located outside the sample (not shown in \autoref{fig:ResultClampedBoundary}(c)); the physical interpretation of this virtual energy creates some obvious and unpleasant conceptual problems. Overall the response is still too soft. Surface correcting the side surfaces improves the situation only marginally, the results are not shown in \autoref{fig:ResultClampedBoundary} since the elastic energy patterns are by eye almost indistinguishable from those of \autoref{fig:ResultClampedBoundary}(c). Uncorrected peridynamics produces a sad result. Elastic energy is strongly concentrated in the nodes directly at the constrained top and bottom surfaces, see the thin red lines at the bottom and top of \autoref{fig:ResultClampedBoundary}(d). These nodes are, in an uncorrected PD simulation, very weakly bonded to the rest of the system and therefore their weak bonds become overstretched, leading to a high energy concentration while the rest of the system is unloaded and the overall stress goes down as seen in \autoref{tab:ResultClampedBoundary}. Fhis feature has undesirable and unphysical consequences when considering fracture based on a bond-stretch criterion, as mode-I cracks will always localize at the top or bottom surface. (Note that, in the surface corrected system, the energy concentration at the top and bottom surfaces is {\em not} associated with any unphysical bond stretch, since the bonds associated with the surface nodes are strongly stiffened by the correction factors). 

\subsection{Indentation}

Moving away from small-displacement situations, we consider the indentation of a square block by a circular indenter. The edge length of the block is taken to be \SI{40}{\mm}, the indenter radius is \SI{15}{\mm}, and the indentation depth is ramped up from \SIrange{0}{2}{\mm}. For FEM discretization we use a grid of square elements of edge length \SI{0.25}{\mm}, and for PD a matching grid of nodes. The horizon is taken to be \SI{1.5}{\mm}, hence $m=1/6$ as in previous simulations. This implies that the indent radius is of the order of the horizon or smaller. 

The indenter is moved downward rigidly, and parts of the surface that get in contact with the indenter move without slip along with the indenter surface. A similar boundary condition is imposed in the PD simulations: nodes that get into contact with the indenter again move without slip along with the indenter surface. 

\begin{figure}[tbh]
	\centering
	\hbox{}
	\includegraphics[width=.6\textwidth]{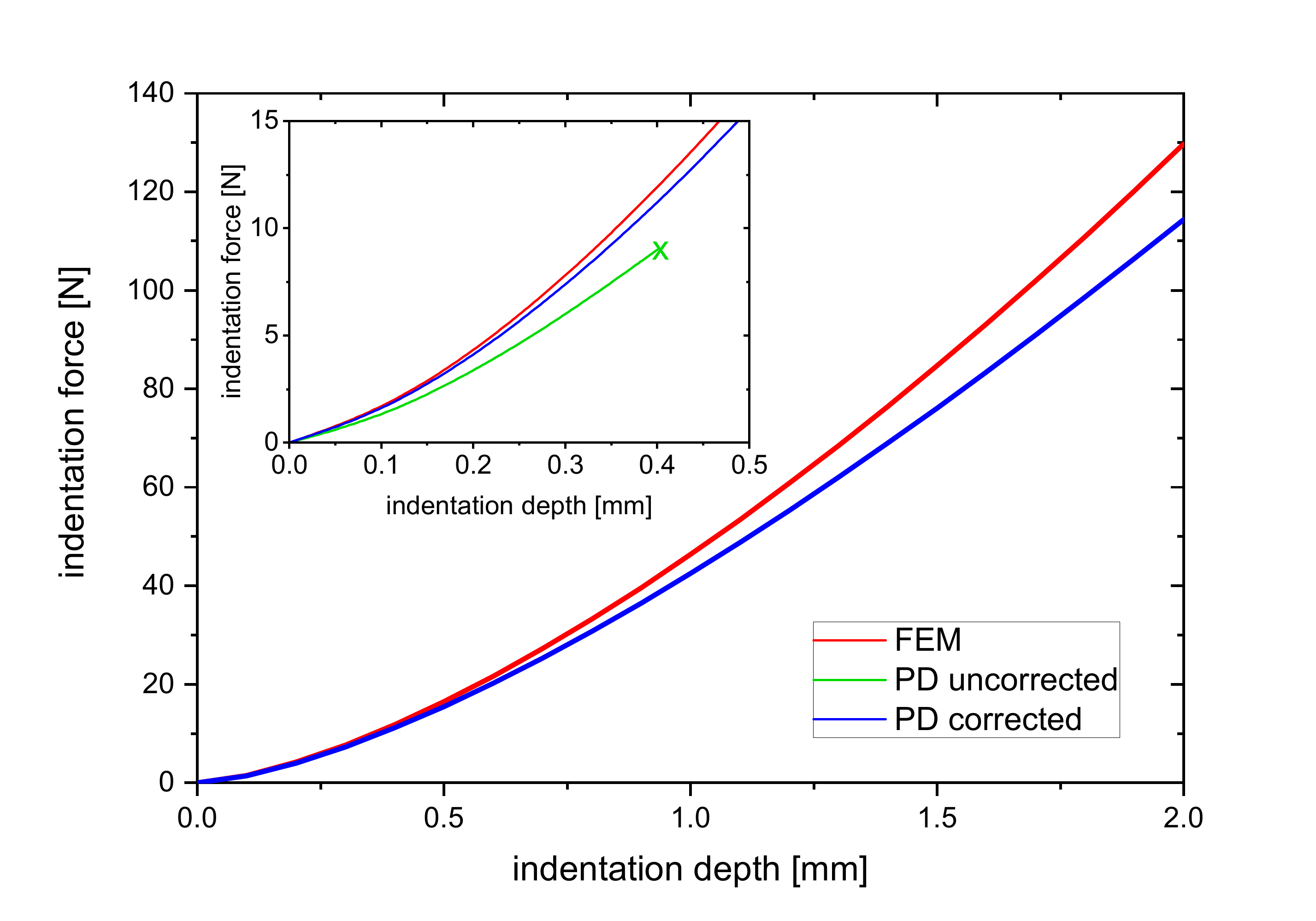}\hfill
	\caption{\label{fig:indentationcurves}
		Simulated indentation curves, FEM reference and corrected PD; inset: initial stage of indentation showing also the curve for uncorrected PD, the green cross indicates the point when the uncorrected PD scheme fails to converge due to loss of uniqueness of the solution.}
\end{figure}

Indentation curves are shown in \autoref{fig:indentationcurves} which compares the results of uncorrected PD, corrected PD and FEM reference. As expected, the indentation curve from the uncorrected PD simulation is significantly softer than the FEM reference, with an indentation force that is about \si{30}{\%} smaller than the FEM reference values. The surface correction reduces this discrepancy to about \SI{6}{\%} (\autoref{fig:indentationcurves}). More importantly, however, the displacement patterns resulting from the uncorrected PD are completely unrealistic, to the extent that, above an indentation depth of about \SI{0.4}{mm}, the surface nodes in contact with the indenter are pushed {\em beyond} the next row of nodes that was originally at a depth of \SI{0.25}{mm} below the surface. This leads to an inversion of bond direction and consequentially the solver fails to converge (green cross in the inset of \autoref{fig:indentationcurves}). This is not a mere problem of numerics: At this point, the material manifold becomes multiple-valued and as a result the problem becomes both physically meaningless and mathematically undefined. 

\begin{figure}[tbh]
	\centering
	\hbox{}
	\includegraphics[width=\textwidth]{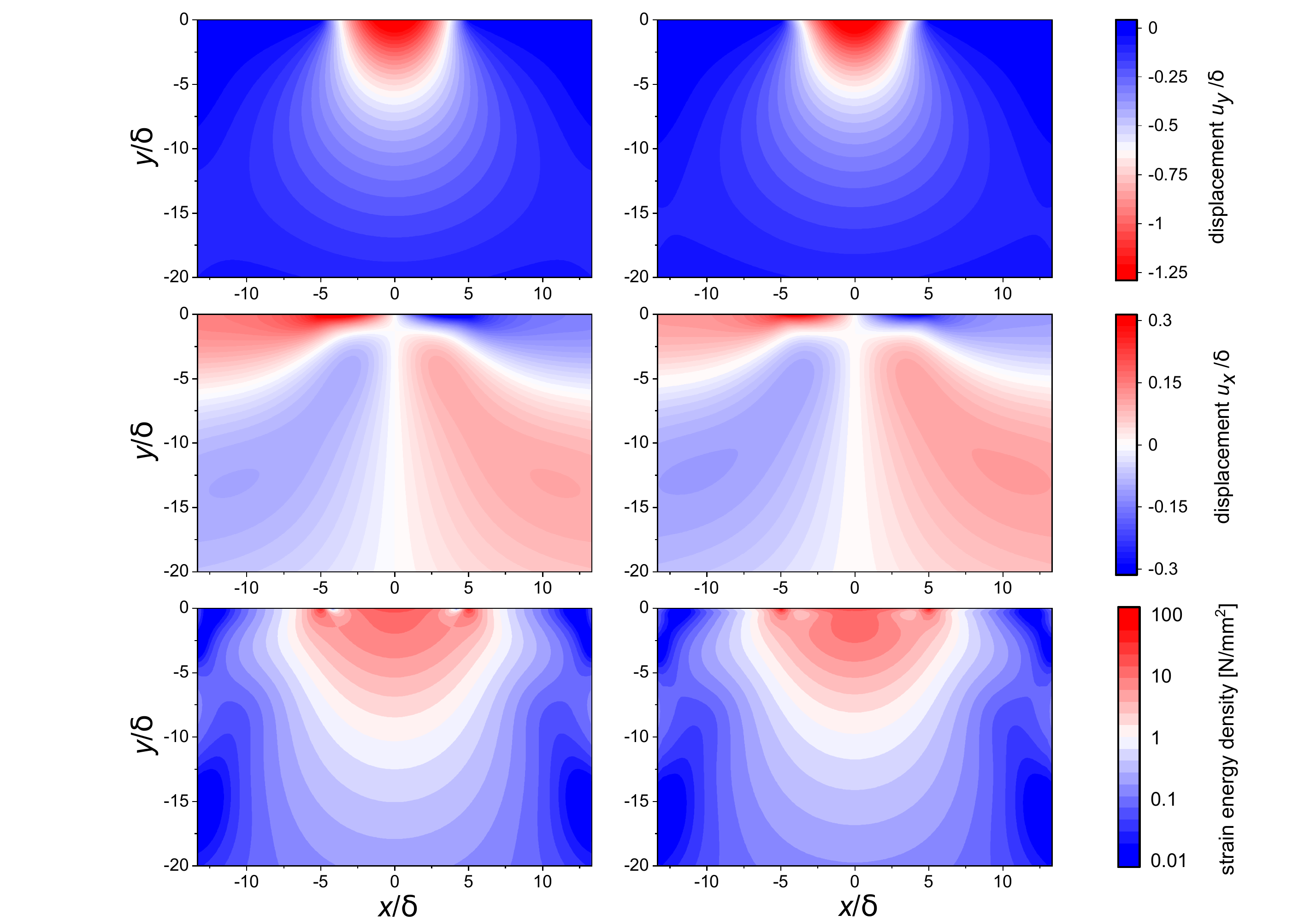}\hfill
	\caption{\label{fig:indentationcomparison}
		Spatial patterns of displacement and strain energy density for simulated indentation; left graphs: FEM reference, right graphs: surface corrected PD; top row: displacement in $y$ direction (direction of motion of the indenter), center row: displacement in $x$ direction, bottom row: strain energy.}
\end{figure}

On the other hand, the corrected PD scheme is doing an excellent job at reproducing both the displacement fields and energy densities obtained from the FEM reference calculation. This is illustrated in \autoref{fig:indentationcomparison} which demonstrates near-perfect agreement between the displacement fields and strain energy density patterns in FEM and corrected PD simulations. In particular, the corrected PD captures correctly the large displacement gradients and high energy densities directly underneath the indenter.

\section{Discussion and Conclusions}
\label{sec:5}

The presented method allows to correct the peridynamic surface effect and to match the behavior of a classical continuum even in situations where the horizon radius is comparable with relevant  dimensions of the sample, or characteristic lengths of the deformation problem. A particular advantage is that the scheme works well in situations where boundary displacements or forces are directly and exclusively applied to boundary nodes. This allows to directly and intuitively transfer boundary value problems from a finite element to a peridynamic setting. 

The superiority of the present approach relative to existing energy based surface correction schemes lies in the directionality of the bond stiffening rule. The presence of the surface breaks the isotropic symmetry of the bulk material and this reduced symmetry should be reflected by the correction factors. Previous attempts to formulate energy based surface corrections in terms of correction factors that are constant over the entire horizon cannot adequately capture the displacement patterns under different surface loadings, whereas the present scheme does an excellent job in this respect. 

The ability of the present formalism to accurately match the displacement fields of the classical reference continuum is particularly useful when using bond failure criteria that are based on bond stretch. Similarly, the accuracy in reproducing elastic energy densities may allow the formulation of energy-based failure criteria. Further work is, however, required to extend the present bond correction scheme to situations where damage by multiple bond failures is present, since such situations have no direct correspondence in the classical reference continuum. 

\section*{Declarations}
\subsection*{Competing interests}
  The authors declare that they have no competing interests.
  
\subsection*{Author's contributions}
J.R. performed peridynamic simulations and data analysis, S.S. performed FEM reference calculations, M.Z. devised the surface correction method with support of P.S. and drafted the manuscript. The manuscript was edited and approved jointly by all authors. 

\subsection*{Funding}
This work was funded by the Deutsche Forschungsgemeinschaft (DFG, German Research Foundation) - 377472739/GRK 2423/1-2019. The authors gratefully acknowledge this support. 

\subsection*{Acknowledgements}
.
\subsection*{Availability of data and material}
Not applicable
%%  After submission of the .TEX file,                     %%
%%  you will be prompted to submit your .BBL file.         %%

% if your bibliography is in bibtex format, use those commands:
\bibliographystyle{elsarticle-num} % Style BST file (bmc-mathphys, vancouver, spbasic).
\bibliography{periref}   

\newpage

\appendix
\setcounter{equation}{0}
\setcounter{figure}{0}
\setcounter{table}{0}
\renewcommand{\theequation}{\Alph{section}.\arabic{equation}}

\end{document}